\documentclass{cjpsuppl}% for LaTeX 2e
\usepackage{graphicx}
\begin{document}
\title{Ultraperipheral Collisions}
\authori{Kai Hencken}
\addressi{Institut f\"ur Physik, Universit\"at Basel, Basel, Switzerland}
\authorii{Gerhard Baur}
\addressii{Institut f\"ur Kernphysik, Forschungszentrum J\"ulich,
  J\"ulich, Germany}
\authoriii{Ute Dreyer, Dirk Trautmann}   
\addressiii{Institut f\"ur Physik, Universit\"at Basel, Basel, Switzerland}
\authoriv{}     \addressiv{}
%\authoriv{Dirk Trautmann}    
%\addressiv{Institut f\"ur Physik, Universit\"at Basel, Basel, Switzerland}
\authorv{}     \addressv{}
\authorvi{}    \addressvi{}
\headtitle{Ultraperipheral Collisions}
\headauthor{Kai Hencken}
\lastevenhead{Kai Hencken: Ultraperipheral Collisions}
%%%%%%%%%%%%%% Pro editory supplementu: %%%%%%%%%%%%%%%
\refnum{}%slouzi editorum pro evidenci; nakonec {}
\daterec{15 September 2004;\\final version 15 September 2004}
\suppl{A}  \year{2004} \setcounter{page}{21}
\firstpage{1}
%\lastpage{000}
%\makefirsttitle
%%%%%%%%%%%%%%%%%%%%%%%%%%%%%%%%%%%%%%%%%%%%%%
\maketitle

\begin{abstract}
Ultraperipheral collisions at heavy ion colliders use the strong
Coulomb fields surrounding the ions to study photon-photon and
photon-hadron processes at high energy. A number of 
processes of interest are discussed here.
\end{abstract}

\section{Introduction}
\begin{figure}[!t]
  \centering
  \begin{minipage}[b]{4.5cm}
\resizebox{4cm}{!}{\includegraphics{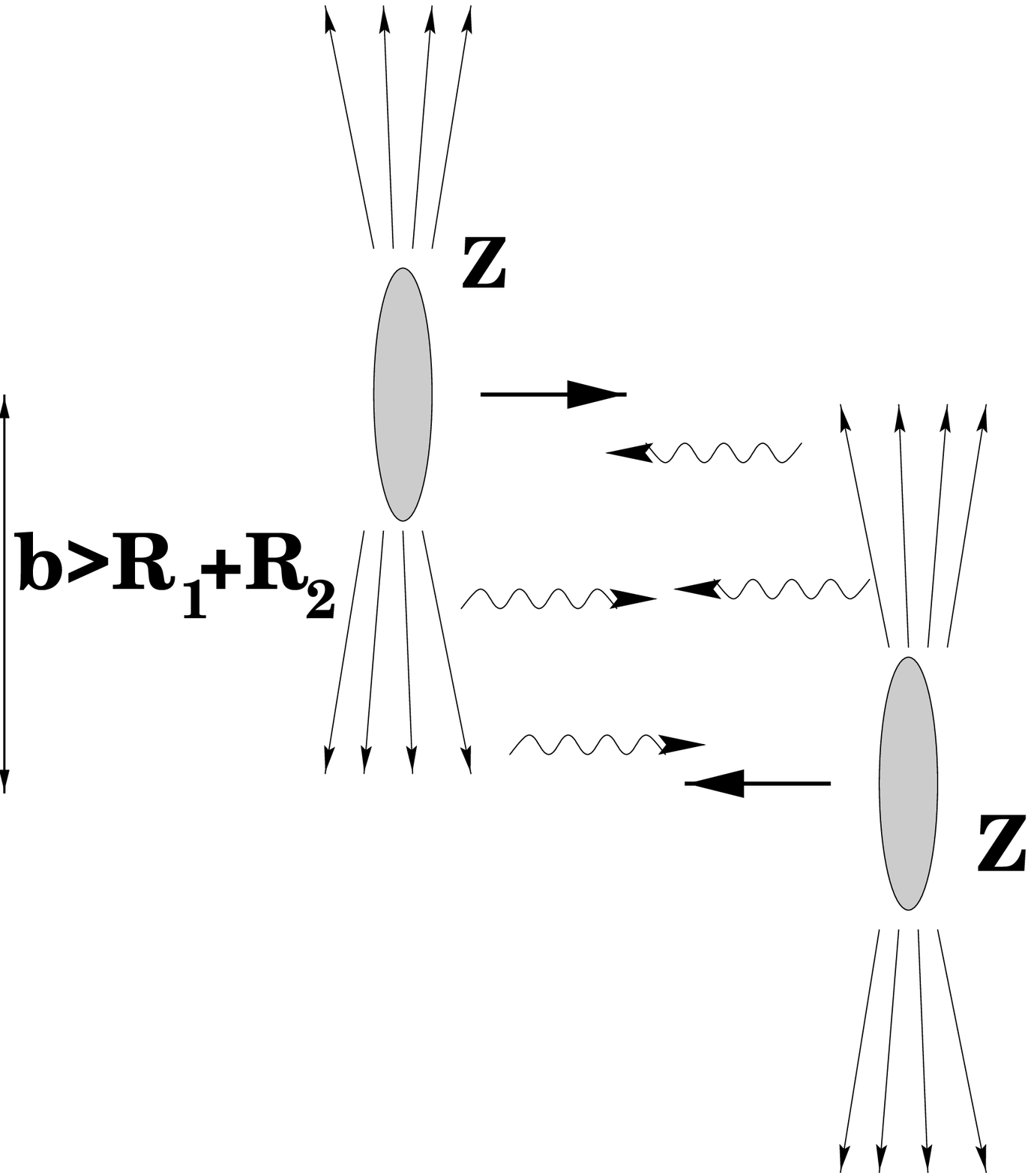}}
  \end{minipage}
~~~
 \begin{minipage}[b]{4cm}
\resizebox{1.75cm}{!}{\includegraphics{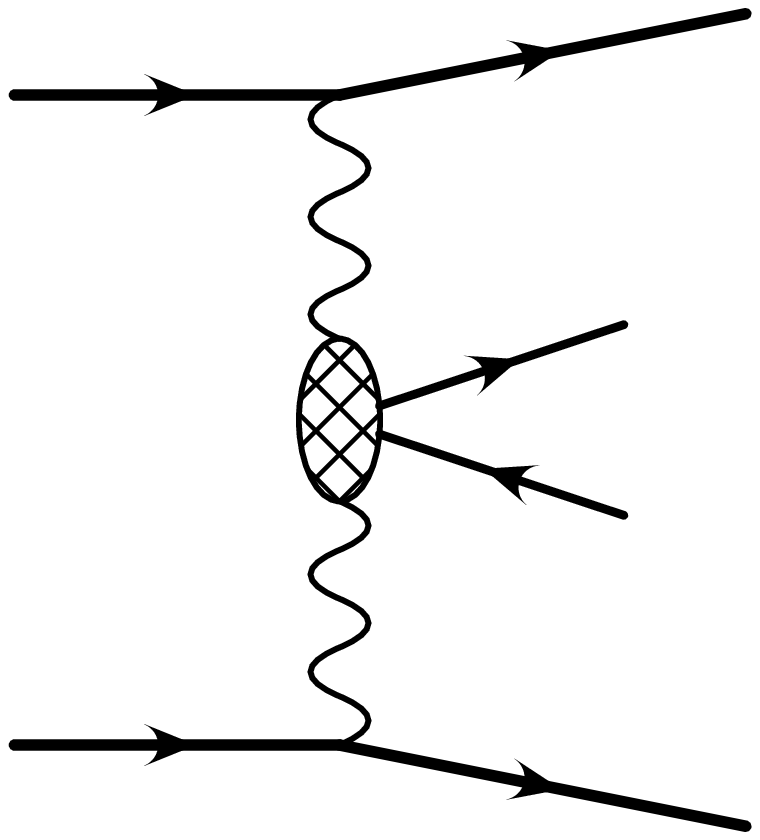}}\\
~\\
\resizebox{1.75cm}{!}{\includegraphics{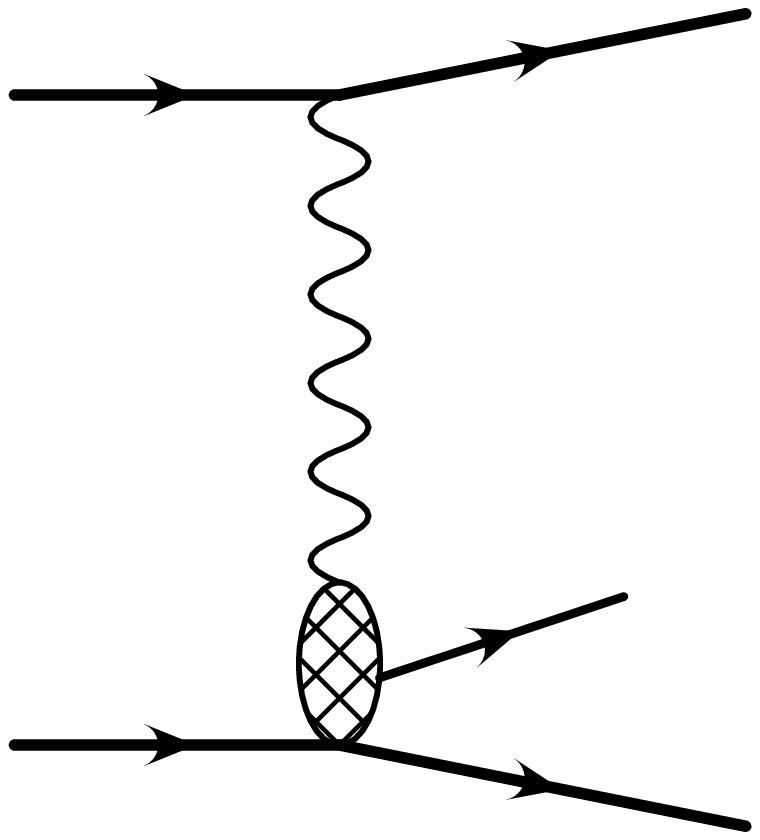}}
\end{minipage}
  \caption{The Coulomb field surrounding the heavy ions in
 relativistic heavy ion collisions can be seen as a flux of quasireal
 equivalent photons. In ultraperipheral collisions (UPC) they are used
 for photon-photon and photon-nucleus processes.}
  \label{fig:UPC}
\end{figure}

Ultraperipheral collisions (UPC) are the processes that occur at
impact parameter $b>2R$, when the two ions do not interact hadronically.
Instead one uses the strong Coulomb field
surrounding the ions for
elementary particle processes, see Fig.~\ref{fig:UPC}. 
The coherence of all the protons in the ion leads to an
enhancement factor of $Z^4$ ($\gamma\gamma$) or $Z^2$ ($\gamma A$), 
respectively, compared to the one for $pp$ or $ee$.
This offers the possibility to study a number of interesting
$\gamma\gamma$ and $\gamma A$ processes at energies and masses, which
were up to now not available. The flux of ``equivalent photons'' (quasireal
photons)
goes up to the region of 100~GeV for $\gamma\gamma$ collisions 
(collider rest frame) and up to 500~TeV for $\gamma A$
collisions (rest frame of the target ion).

Ultraperipheral collisions have been studied for some time now 
\cite{BaurHTS02,BaurHT98,KraussGS97,BertulaniB88} and are part of the
heavy ion program of ALICE \cite{ALICEPPR}, ATLAS \cite{ATLASHI}
and CMS \cite{CMSHI,CMS-REVIEW}. 
Detailed studies have been made already for some
specific processes for different LHC detectors, 
as will be discussed in the following.

Photon-photon physics has been studied in $ee$ collisions at
LEP at CERN, photon-proton, photon-photon and also photon-ion
collisions were studied at HERA at DESY.
UPC allow to extend these successful studies to higher energies and higher
luminosities. The main theoretical tool is the ``equivalent photons 
approximation'' or ``Fermi-Weizs\"acker-Williams
method'', first developed by Fermi \cite{Fermi24} and extended to
relativistic energies by Weizs\"acker and Williams
\cite{Williams34}. Its application was studied
in detail in connection with lepton colliders in \cite{BudnevGM75}.

For the ion case there are some important differences compared to the
lepton case, which need to be taken into account: The nucleus is not a 
point-like object, but has a finite size, described by its elastic form factor
$F(k^2)$. As $F(k^2)\approx 1$ only for $k^2< \frac{1}{R^2}$, with $R$
the nuclear radius, this form factor leads to two restrictions; on the one hand
the transverse momentum is limited to $k_\perp< 1/R \approx
30$~MeV. This means that in contrast to lepton beams, only
``quasireal'' photons can be studied. On the other hand it also
limits the maximum energy which can be taken by the photon to $\omega
< \gamma/R$. This maximum energy corresponds only to a tiny fraction
of the total energy of the ion
\begin{equation}
  x_{max}=\frac{\omega_{max}}{E_{ion}}\approx\frac{1}{R M_N A} = 
\frac{\lambda_C(A)}{R}
\end{equation}
One finds $4\times 10^{-3}$ for O, $1.4\times 10^{-4}$ for Pb.

Furthermore the ions are
interacting hadronically if their impact parameter is closer than $2R$. Whereas
photon-photon and photon-nucleus processes will still take place at these
collisions, they are completely covered by the hadronic
processes. Therefore these collisions need
to be removed to get the usable photon-photon luminosity.

Finally due to the strong fields there is also the possibility of
additional photon exchanges and photon excitation processes. 
Whereas they were first seen as a nuisance \cite{BaltzS98}, as most
of them lead to breakup, e.g., neutron emission from the ions, and therefore
the clean ``no breakup'' condition would be spoiled, they have now
found some interesting applications, see below.

In order to describe the equivalent photons the semiclassical approach
was found to be useful, as it allows to take into
account all the ``complication'' discussed above. There exist already
some reviews, where this approach is discussed \cite{KraussGS97,BaurHT98,BaurHTS02}.

\section{Equivalent photon spectra and luminosities at the LHC}

Integrating over all allowed impact parameter, one gets the effective 
photon-photon luminosity for the production of a final state with
invariant mass $W$ and with rapidity $Y$ in the semiclassical picture as
\begin{eqnarray}
\frac{dL_{\gamma\gamma}}{dWdY} &=& \frac{2}{W} 
\int_{R_{min}} d^2b_1 \int_{R_{min}} d^2b_2 \nonumber\\
& &\times N_1(\frac{W}{2}e^Y,b_1) N_2(\frac{W}{2}e^{-Y},b_2) 
\Theta(|\vec b_1 + \vec b_2| - R_{min}).
\end{eqnarray}
Here $N_1$ and $N_2$ are the impact parameter dependent equivalent
photon numbers. For further details, see \cite{BaurHTS02}.
With this the cross section for a $\gamma\gamma$ process 
factorizes into a luminosity and a (real) elementary $\gamma\gamma$
cross section
\begin{equation}
\sigma(A+A\rightarrow A+A+X) = \int dWdY \frac{dL_{\gamma\gamma}}{dWdY} \sigma(\gamma+\gamma \rightarrow X, W).
\end{equation}

For photonuclear reactions the equivalent photon number is simply
given by integrating over the allowed impact parameter between the two ions
\begin{equation}
n(\omega) = \int_{R_{min}}^\infty 2 \pi b db N(\omega,b).  
\end{equation}
and we get the cross section as
\begin{equation}
\sigma(AA \rightarrow A+X) = \int d\omega n(\omega) \sigma(\gamma A\rightarrow X, \omega)
\end{equation}
 
\begin{figure}[!t]
\begin{center}
\resizebox{6cm}{!}{\includegraphics{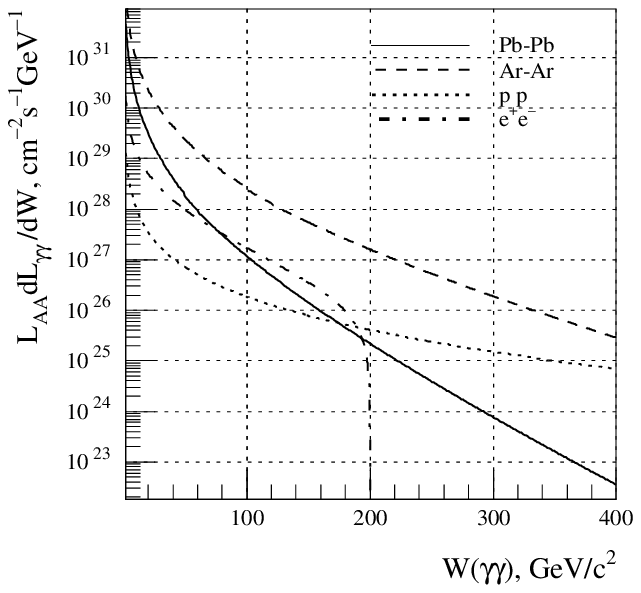}}(a)~~~
\resizebox{5.5cm}{!}{\includegraphics{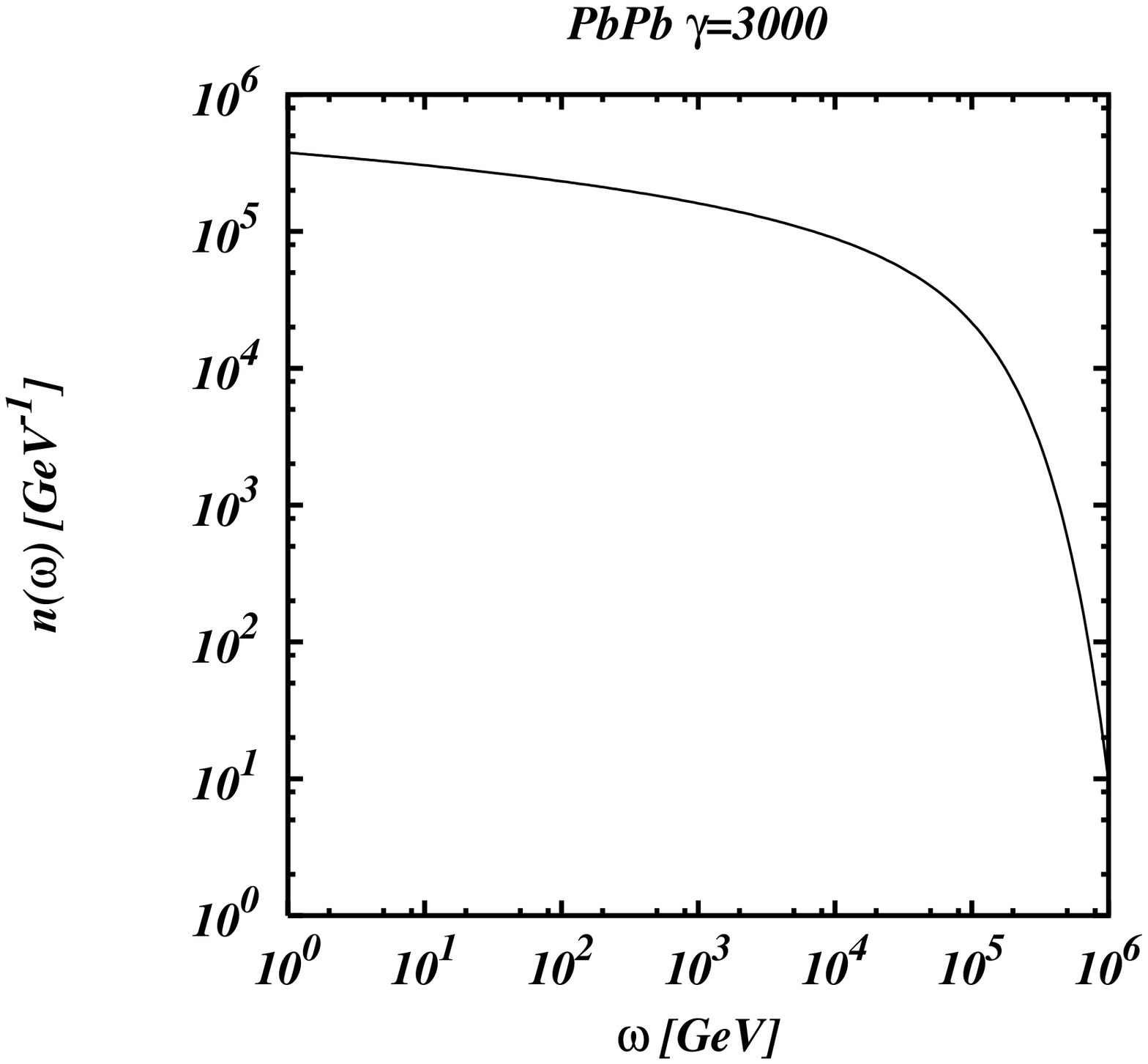}}(b)
\end{center}
\caption{(a)Effective photon-photon luminosities for different ion species. 
(b) Equivalent Photon number for Pb\ Pb collisions at the LHC.}
  \label{fig:gglum}
\end{figure}

Taking into account the expected different ion-ion luminosities for the
LHC, see Table~\ref{tab:lumparam}, one gets the effective
photon-photon luminosity as shown in Fig.~\ref{fig:gglum}, taken from \cite{BaurHTS02}. The available invariant masses for the
$\gamma\gamma$ system are beyond what has been achieved at LEP. This
figure also shows that $pp$ or ArAr collisions seem to be more favorable
compared to PbPb. This is due to the fact, that the PbPb beam luminosity is
five orders of magnitude smaller than the $pp$ luminosity, which
compensates the $Z^4$ enhancement \cite{BrandtEM94b}. Also for $pp$
collisions the invariant mass of the $\gamma\gamma$ system goes beyond
what is available for $PbPb$. The reason for this is the smaller size
of the proton, giving rise to a harder photon spectrum.
One might conclude from this that there is no
real advantage in using heavy ion beams for $\gamma\gamma$ collisions,
but one should keep in mind, that the most important background are 
diffractive processes (``Pomeron-Pomeron'' processes) \cite{EngelRR97}.
These events can
have the same characteristic of leaving the two ions intact,
which is an essential signal to distinguish ultraperipheral collisions
from, e.g., grazing collisions. The coherence of the photon emission
helps to reduce this background. In the case of $pp$ 
collisions diffractive processes clearly dominate the
electromagnetic ones and additional care needs to be taken to
distinguish the two. The electromagnetic cross section grows
like $Z^4/Z^2$. The coherent diffractive 
processes is sensitive to the surface region of the two ion, the cross 
section is therefore proportional to
$A^\delta$ with $\delta\approx 1/3$ \cite{EngelRR97}. 
For lead ions the
electromagnetic processes are expected to be dominant \cite{Felix97,FELIX}.

\begin{table}[!t]
  \centering
  \begin{tabular}{|c|c|}
\hline
     & $L_{AA}$ \\
\hline
  p\ p & $1.4\times 10^{31}$ cm$^{-2}$ s$^{-1}$ \\
  Ar\ Ar & $5.2\times 10^{29}$ cm$^{-2}$ s$^{-1}$\\
  Pb\ Pb & $4.2\times 10^{26}$ cm$^{-2}$ s$^{-1}$\\
\hline
  \end{tabular}
  \caption{Beam Luminosities for different ions species at the LHC}
  \label{tab:lumparam}
\end{table}

For photonuclear reactions one takes the photon energy in the
rest frame of the ion, using 
$\gamma_{ion}=2\gamma_{coll}^2-1$ instead of $\gamma_{coll}$. This
leads to photon energies of up to 500~TeV, way beyond the
possibilities of HERA, see Fig.~\ref{fig:gglum}(b). 
In addition the use of heavy ions is an
advantage here as well, photonuclear processes on ions 
can be measured at these high energies.

\section{Potential for $\gamma\gamma$ physics}

% Eventuell nach unten
\subsection{New physics searches}
The production of the Higgs boson was studied by a number of people
in the past, see \cite{KraussGS97} for a detailed review. It was found
that electromagnetic production is favorable compared to hadronic
production, as it allows for rather clean events
\cite{DreesGN94,OhnemusWZ94}. Unfortunately due to
the lower effective luminosity and the higher mass limit from the LEP Higgs 
searches, rates for a SM or MSSM Higgs are rather
small \cite{BaurHTS98}. 
This will make UPC for a search or the study of the
Higgs boson very unfavorable. Still models with a larger coupling of
the Higgs to two photons, e.g., a ``leptophobic'' Higgs, or
nonstandard models with a light Higgs might still be possible and
could be investigated in this way \cite{Krawczyk96,ChoudhuryK97,LiettiNRR01}.

Other particles have also been studied to be detected in photon-photon 
processes: With the increased mass limit, the production rates of SUSY
particles were again found to be to small \cite{HenckenKKS96,BaurHTS02}. 
Magnetic monopoles have been searched by
looking for $\gamma\gamma\rightarrow\gamma\gamma$ with a large
transverse momentum at the TEVATRON \cite{GinzburgS98,Abbott98}. 
Due to the strong coupling of the magnetic monopole to the photon this
cross section is strongly enhanced. Such a search
could also be feasible at the LHC and could increase the mass limit.
The process $\gamma\gamma\rightarrow\gamma\gamma$ was also proposed to
be used to study the $\sigma$ meson \cite{RoldaoN00}.

\subsection{Tagging of the final protons}

At CMS/TOTEM one has the possibility to detect protons, which have
lost more than about 1\% of their energy
\cite{Piotrzkowski00,Piotrzkowskihere}, corresponding to a photon energy
of 70~GeV. This opens the possibility to
study $\gamma\gamma$ processes at high energies. It also allows to
determine directly the energy of the emitted photon and therefore the
mass of the $\gamma\gamma$ system. Loosening the restriction of
detecting both ions allows to increase the luminosity and also extend
the invariant mass spectrum to lower energies. This allows to study 
electromagnetic processes in the electroweak sector. In $pA$
collisions one can look also for $\gamma\gamma$, $\gamma p$ or
$\gamma A$ processes. Due to the small $x_{max}$ it will not be possible to 
detect the lead ions in this case.

Diffractive processes will again be a background. As the Pomeron also
has a high probability 
to be emitted with the
proton remaining intact, they cannot be distinguished from the event
characteristics from UPC. On the other hand the photon has a very
narrow transverse momentum distribution, whereas the Pomeron leads to
a momentum distribution of the proton in the area of several 100~MeV. There is
a theoretical limitation for using $pp$ beams for UPC, coming from the
occurance of overlapping events \cite{DreesGN94}. E.g., at the high
luminosity run of the LHC with $L_{pp} \sim 10^{34}$~cm$^{-2}$s$^{-1}$
there will always be hadronic interaction in each bunch crossing,
making the study of UPC rather difficult.

\subsection{$\gamma\gamma$ processes at lower energies}
Even at lower invariant masses there are a number of processes of
interest. One such possibility is double vector meson production,
which was already studied in connection with FELIX \cite{Felix97,FELIX}. It
allows for a test of the soft factorization hypothesis
\begin{equation}
\frac{\sigma(\gamma p\rightarrow V_1 p)}{dt}
\frac{\sigma(\gamma p\rightarrow V_2 p)}{dt}
=
\frac{\sigma(\gamma \gamma \rightarrow V_1 V_2)}{dt}
\frac{\sigma(pp\rightarrow pp)}{dt}
\end{equation}
as well as
\begin{equation}
\frac{\sigma(\gamma \gamma \rightarrow V_1 V_1)}{dt}
\frac{\sigma(\gamma \gamma \rightarrow V_2 V_2)}{dt}
=
\left(\frac{\sigma(\gamma \gamma \rightarrow V_1 V_2)}{dt}\right)^2,
\end{equation}
which allows to relate this process to other vector meson production
processes. Deviations from this factorization are expected to be
large.

\begin{figure}[!t]
  \centering
  \resizebox{4.75cm}{!}{\includegraphics{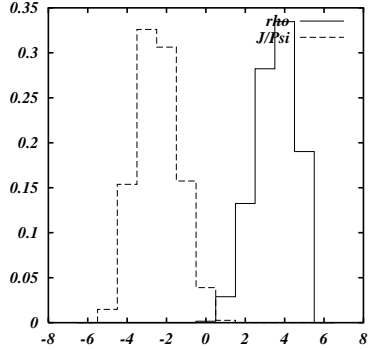}}
  \caption{Unsymmetric distribution of the two vector mesons from $A+A
  \rightarrow A+A+\rho+J/\Psi$ as a function of the rapidity. Shown
  are the results for PbPb collisions at the LHC in a Vector Dominance
  model.}
  \label{fig:rapVV}
\end{figure}

Detailed studies taking into account the possibility to measure the muons from
the decay in the ALICE Muon arm are under way. The muons can be used as 
L0 trigger at ALICE. The rapidity
distribution, which centers around $Y=3$, see Fig.~\ref{fig:rapVV} 
agrees quite well with the acceptance region of the Muon Arm of ALICE.

Another process of interest would be $\gamma\gamma\rightarrow$~hadrons,
which has been studied at L3 at LEP
\cite{L3:97,Acciarri00b}. 
Deviation from the Regge universality was found at
high energies and especially for final states containing heavy quarks.
The total cross section measured by all the
detectors at LEP has an overall uncertainty of about 40\% coming
from the uncertainty of different Monte Carlo generators in the
forward region, which was not covered by LEP \cite{L3:01}. 
Here one might hope that
the detectors at the LHC, some of which have extensive forward
detection possibilities, could contribute to this process.
Unfortunately it is not easy to measure such hadronic final states and to
distinguish them from hadronic processes.

Finally meson spectroscopy of lighter mesons, containing $c$ and $b$
quarks can be studied. The $\gamma\gamma$ decay width of these
mesons gives an insight into the question, whether they are
predominantly build from quark or gluon degrees of freedom. QED
processes can be studied in this way too. Lepton pair production is a
process of interest due to its large cross section (about 200~kbarn for
PbPb collisions at the LHC), allowing for a luminosity measurement, 
but also as multiple pair
production occurs in single collisions, an interesting higher order
QED effect, see \cite{Adams04,HenckenBT04} and also \cite{BaurHTS02} 
were further references can be found. Bound-free pair production (also
called ``electron capture from pair production'' ECPP) is the
process, where the electron is not produced as a free particle, but
into the bound state of one of the ions \cite{BaurHTS02}.
Even though this
is only a tiny fraction of the free pair production process, it has a
total cross section of 200~barn for PbPb collisions at the LHC and is
(together with the electromagnetic excitations of the ions discussed
below) the dominant loss process of the lead beam. In addition the
$Pb^{81+}$ ions hit the beam pipe at a very definite point and lead to
the deposition of a large amount of energy \cite{Klein01,Brandt00}. 
The quench limit of the
superconducting magnets gives for PbPb beams the limit for the maximal
beam luminosity. Finally positronium and muonium are produced in
large numbers in both ortho- and para-states and could be studied
as well \cite{BaurHT98,BaurHTS02}.

\begin{figure}[!t]
  \centering
  \resizebox{6cm}{!}{\includegraphics{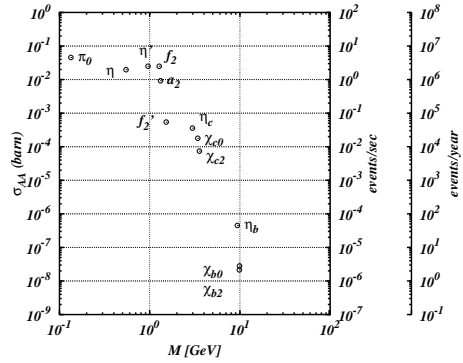}}
  \caption{Cross section and production rates for different mesons in $\gamma\gamma$ collisions for PbPb collisions at the LHC.}
  \label{fig:meson}
\end{figure}

\section{Photon hadron reactions}
\subsection{Photo-excitation of the nucleus}

The photon spectrum available for photonuclear reactions ranges 
up to the TeV regime.
A dominant feature at low energies is the 
excitation of the giant dipole resonance (GDR), followed by other nuclear and nucleon excitations at higher energies. In collisions close to
$b\approx 2R$ the probability to excite one ion is about 75\%, the
excitation of the GDR contributes to this with about 40\%.
The total cross section for this
excitation is about 200~barn \cite{BaronB93,VidovicGS93,Baltz98}. 
As the GDR predominantly deexcites by
neutron emission, this leads to a change in the $Z/A$ ratio of the
ion and it is lost in the beam. Together with the bound free pair
production cross section, mentioned above, it is the main
loss process of the lead ions, limiting the total beam lifetime.
Due to the strong fields, large probabilities
exist for more than one process to occur in one collision. This can
be either the excitation of higher states in the ions (double GDR,
%check ref.
triple GDR) \cite{Pshenichnov00,Pshenichnov01} or of other processes
in connection with a GDR excitation of one or both ions.

\begin{figure}[!t]
  \begin{center}
\resizebox{6cm}{!}{\includegraphics{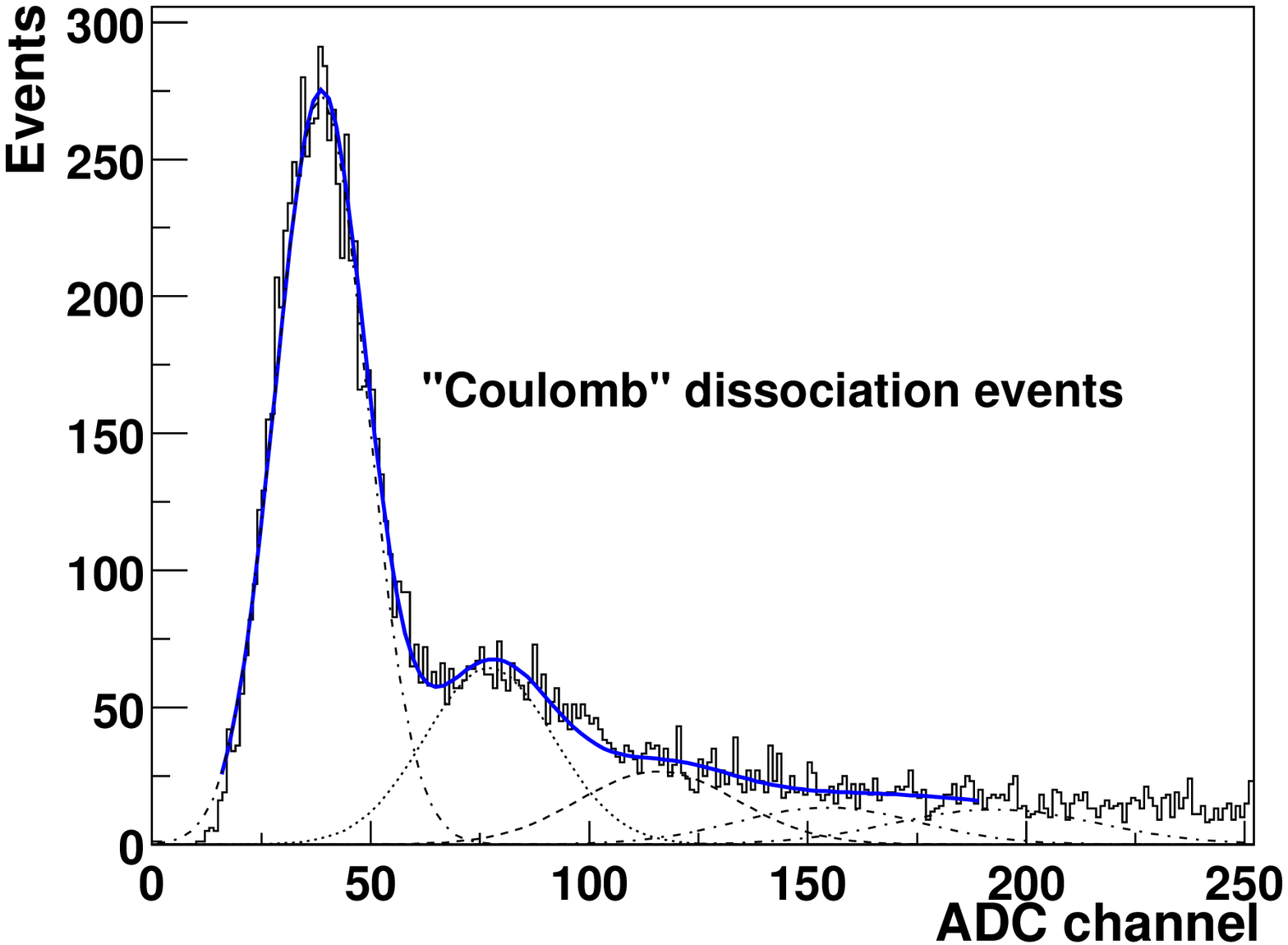}}
\resizebox{6cm}{!}{\includegraphics{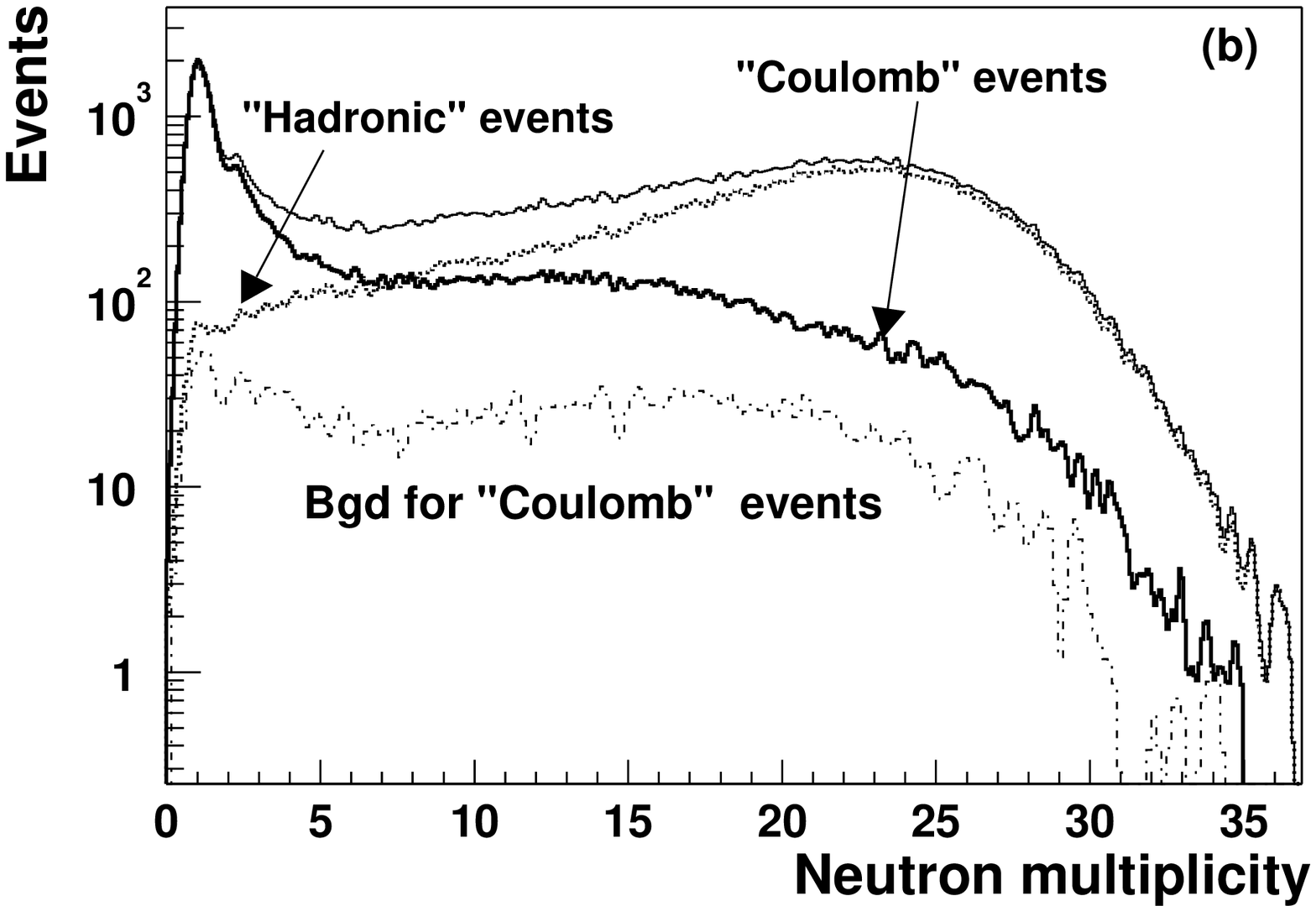}}
\end{center}
  \caption{Deposited energy and neutron multiplicity spectrum for
    Coulomb and hadronic events are shown for the ZDCs at RHIC.}
  \label{fig:mutual}
\end{figure}

The mutual excitation and subsequent emission of the neutrons
from both ions is used at RHIC as a luminosity measurement tool
\cite{Baltz98,White01}, where the neutrons are detected in the ZDCs, 
see Fig.~\ref{fig:mutual}. The measured cross section for different
processes (1,2,$x$ neutron emission) were compared with theoretical
predictions from RELDIS \cite{Pshenichnov98} and good agreement was
found.

These mutual excitations are also a useful trigger for UPCs. Single
neutron detection in one or both ZDCs are a good sign that only an
electromagnetic interaction took place between the ions. This method
was pioneered at STAR/RHIC were vector meson production and also
electron-positron pair production was studied
\cite{Adler02,Adams04,HenckenBT04}.

\subsection{Coherent $\gamma A$ processes}
\begin{figure}[!t]
\begin{center}
\resizebox{8cm}{!}{\includegraphics{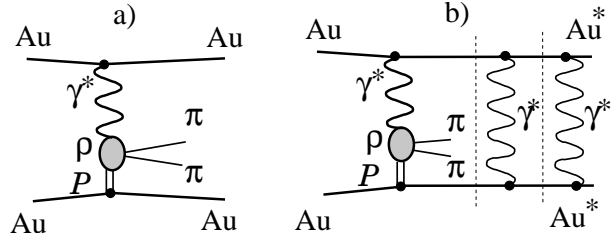}}\\
\end{center}
  \caption{The two measurement of $\rho$-production at STAR: a)
    without and b) with additional electromagnetic excitation of both ions}
  \label{fig:STARtag}
\end{figure}

\begin{figure}[!t]
\begin{center}
\resizebox{!}{4cm}{\includegraphics{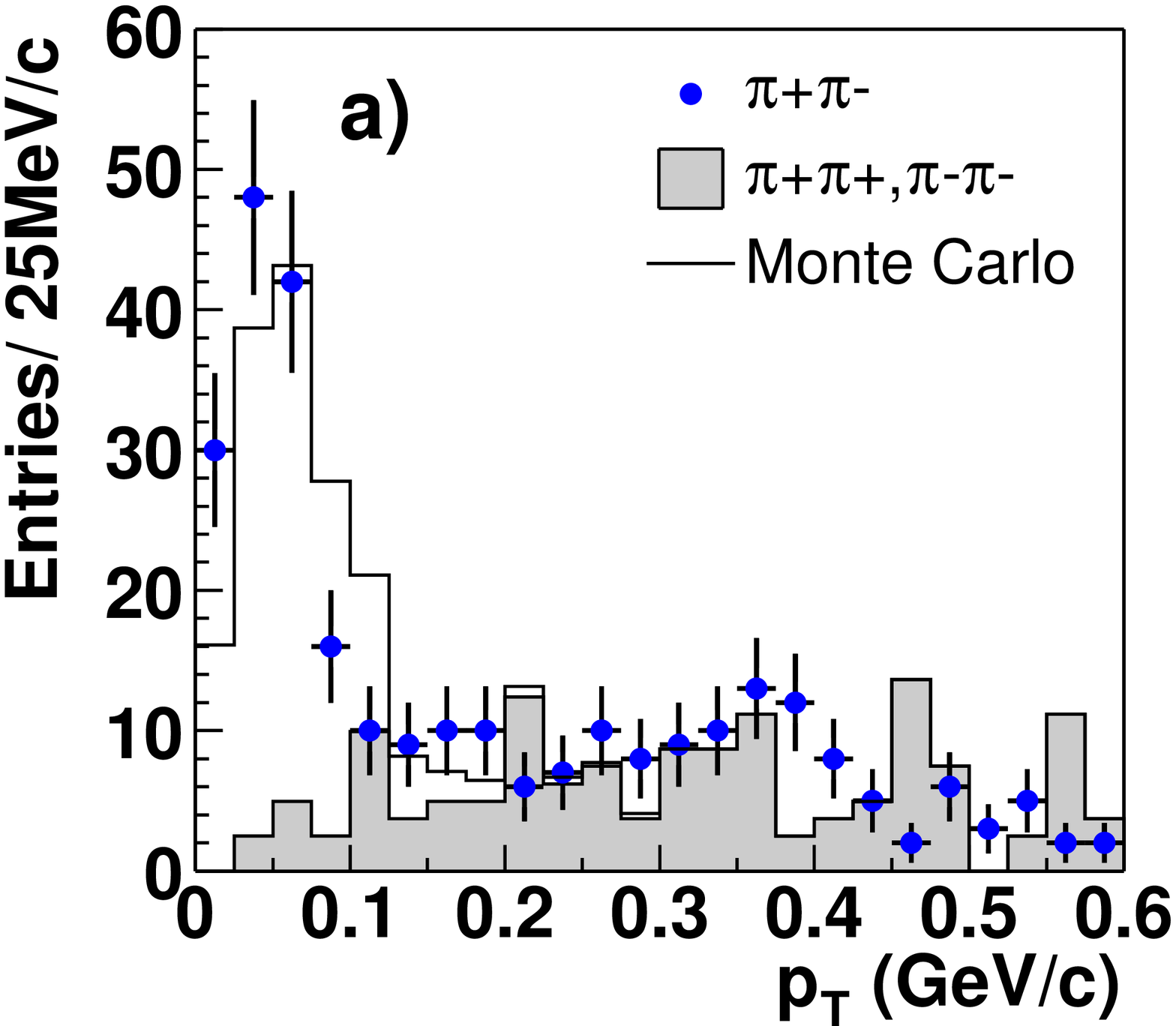}}(a)~~~
\resizebox{!}{4cm}{\includegraphics{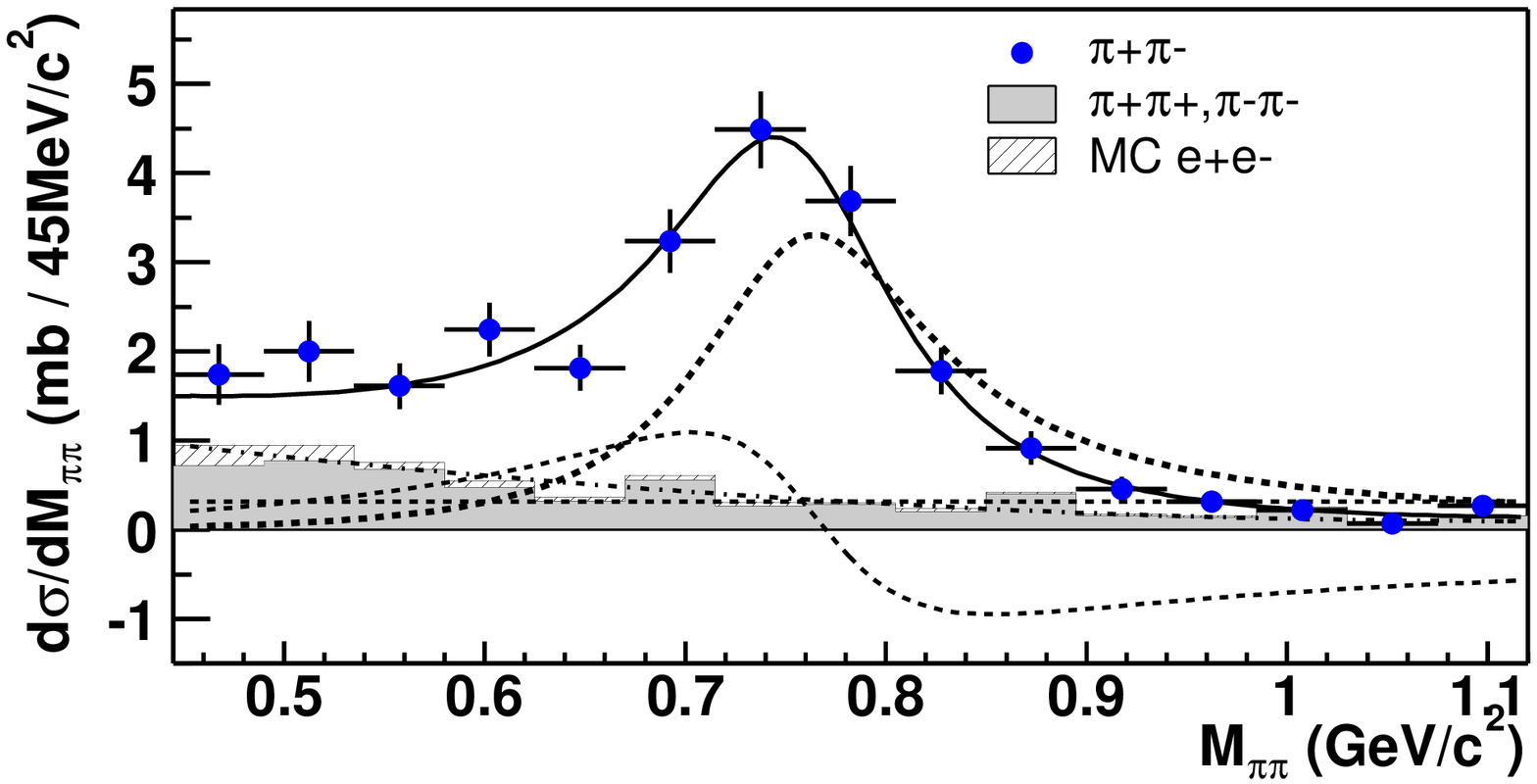}}(b)
\end{center}
  \caption{The transverse momentum (a) and invariant mass spectrum (b)
    for the coherent $\rho$-production at STAR is shown.}
  \label{fig:STARrho}
\end{figure}

At higher energies the diffractive production of vector mesons can be
studied. After theoretical calculations \cite{KleinN99,KleinN00,Baltz:2002pp}
this was for the first time
measured at STAR/RHIC \cite{Adler02,cheunghere,Nystrandhere}. 
In their experiment
the coherent production of $\rho$ mesons was measured with and without
triggering for the additional mutual excitation of the ions, see 
Fig.~\ref{fig:STARtag}. The
coherent production was identified by an enhancement of their
signal at small $p_\perp<1/R$, see Fig.~\ref{fig:STARrho}(a). 
The $\rho$-meson was identified by looking
at the invariant mass spectrum, which could be very nicely fitted to
Breit-Wigner amplitude for the $\rho$-production and a contribution
from direct $\pi^+\pi^-$ production, see Fig.~\ref{fig:STARrho}(b).

There is also an interesting interference phenomenon: As both
ions can act as either the photon source or the target, the two
processes need to be added coherently \cite{KleinN00}. First hints of such an
interference have been seen at STAR \cite{Klein04,Nystrandhere}.


At the LHC these studies can be extended to heavier mesons, especially
to the $J/\Psi$ and even the $\Upsilon$. New phenomena are expected to
occur for these heavier mesons \cite{Frankfurt:2002wc}. Whereas the $\rho$
production is very well reproduced by (Gribov-) Glauber calculations 
\cite{fszstar},
the cross section for $J/\Psi$ and $\Upsilon$ are smaller and are
more sensitive to new phenomena like color transparencies
and nuclear shadowing effects \cite{Frankfurt:2002wc}.
Different predictions have been made
and it is of interest to determine between them at the LHC. 
Simulations have been made for ALICE, using the muons as
L0 trigger.

The additional electromagnetic excitation of the ions can also be helpful 
here: As the additional
excitation restricts the collisions to smaller impact parameter, the
photon spectra is harder than in the unrestricted case \cite{Baur:2003aa}. 
This allows to disentangle the contribution from both ions and to extend the
measurement to rapidities $Y\not=0$ without any model assumption.

\subsection{Photon-gluon fusion processes and quark pdfs}

Also semicoherent processes, where the photon emission occurs
elastically, but there is an incoherent interaction with the target,
are of interest. Inelastic vector meson production is one possible
process of this type.

Most of the interest is focused on photon-gluon
fusion as a possibility to measure the gluon distribution function
inside the nuclei \cite{KraussGS97,KleinNV00,Klein:2002wm}. 
Different models,  predicting nuclear
modifications have been proposed, especially for
small $x$. A precise measurement of these 
``initial state effects''
would also be of importance to model the initial state of ion-ion
collisions in central collisions.

\begin{figure}[!t]
  \begin{center}
\resizebox{2.cm}{!}{\includegraphics{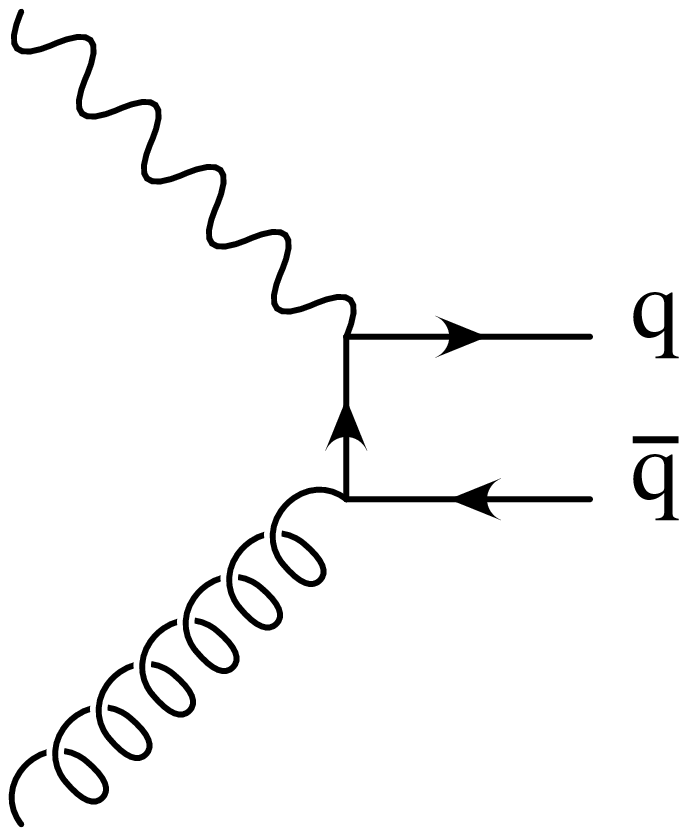}}(a)~~~
\resizebox{2.cm}{!}{\includegraphics{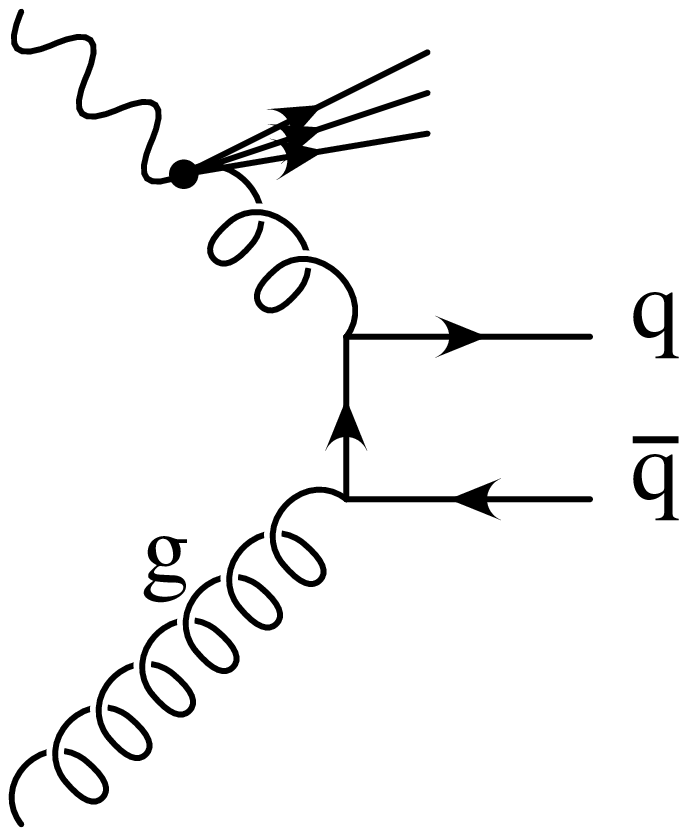}}~~~
\resizebox{2.cm}{!}{\includegraphics{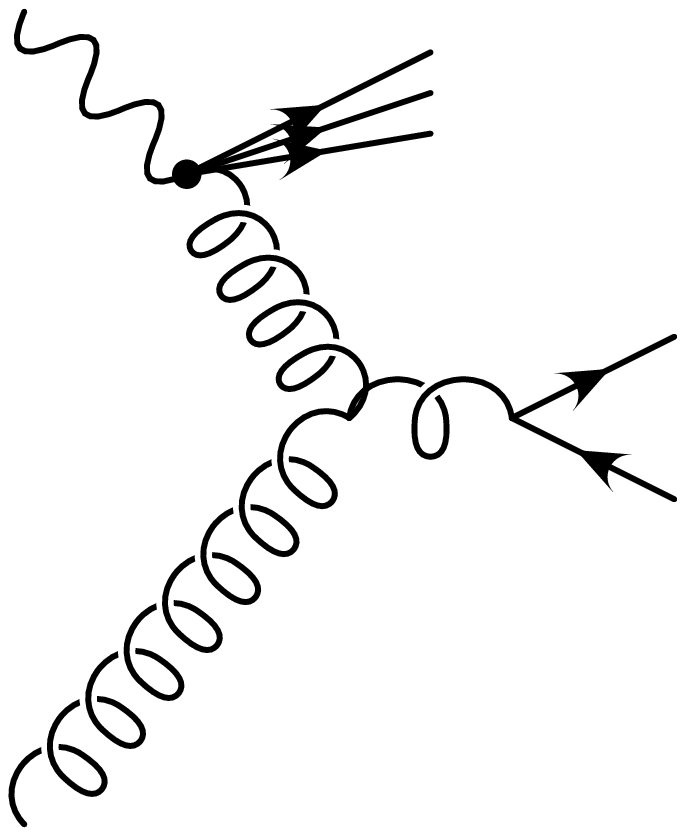}}~~~
\resizebox{2.cm}{!}{\includegraphics{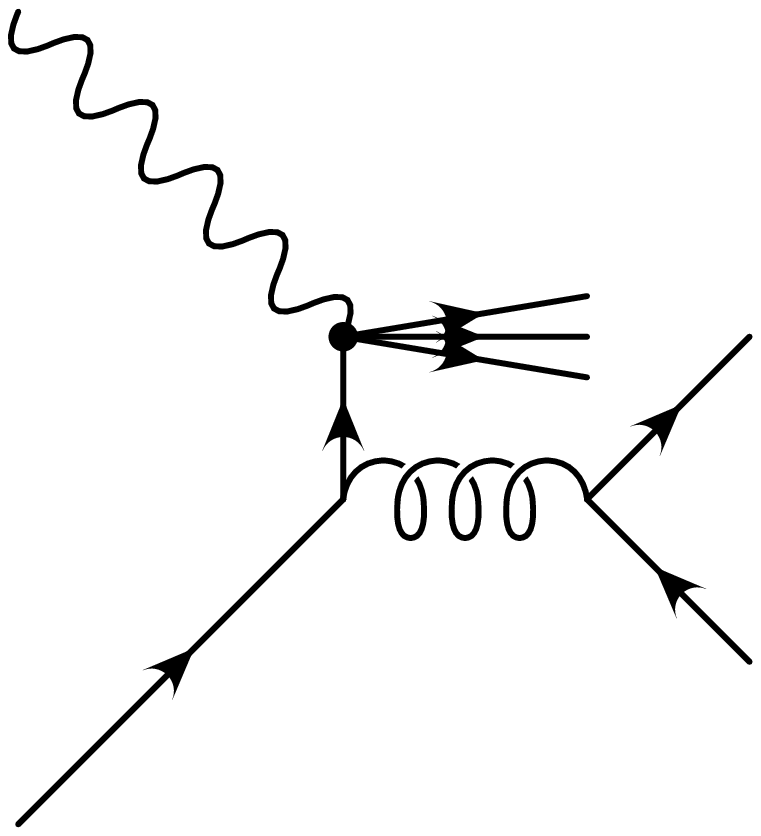}}(b)

\end{center}
  \caption{Feynman diagrams for the photon-gluon fusion process: (a) direct process, (b) resolved processes.}
  \label{fig:ggluon}
\end{figure}

A detailed study of the production of $b$ and $c$ quark pairs was made
for the LHC \cite{Klein:2002wm,Vogt:2004yr} 
taking into account not only the lowest
order diagram, but also resolved contributions, see Fig.~\ref{fig:ggluon}.

\begin{figure}[!t]
  \begin{center}
\resizebox{2.1cm}{!}{\includegraphics{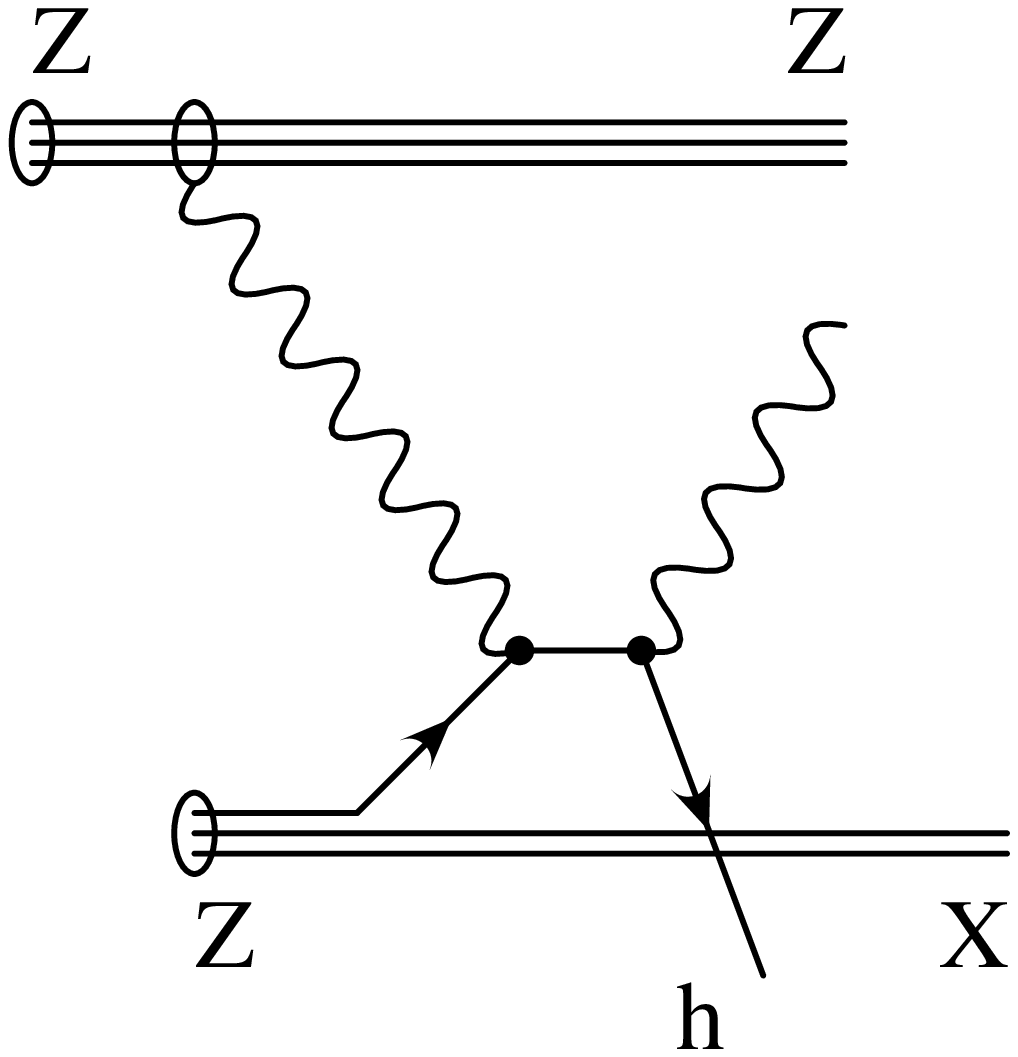}}(a)~~~~~
\resizebox{2.1cm}{!}{\includegraphics{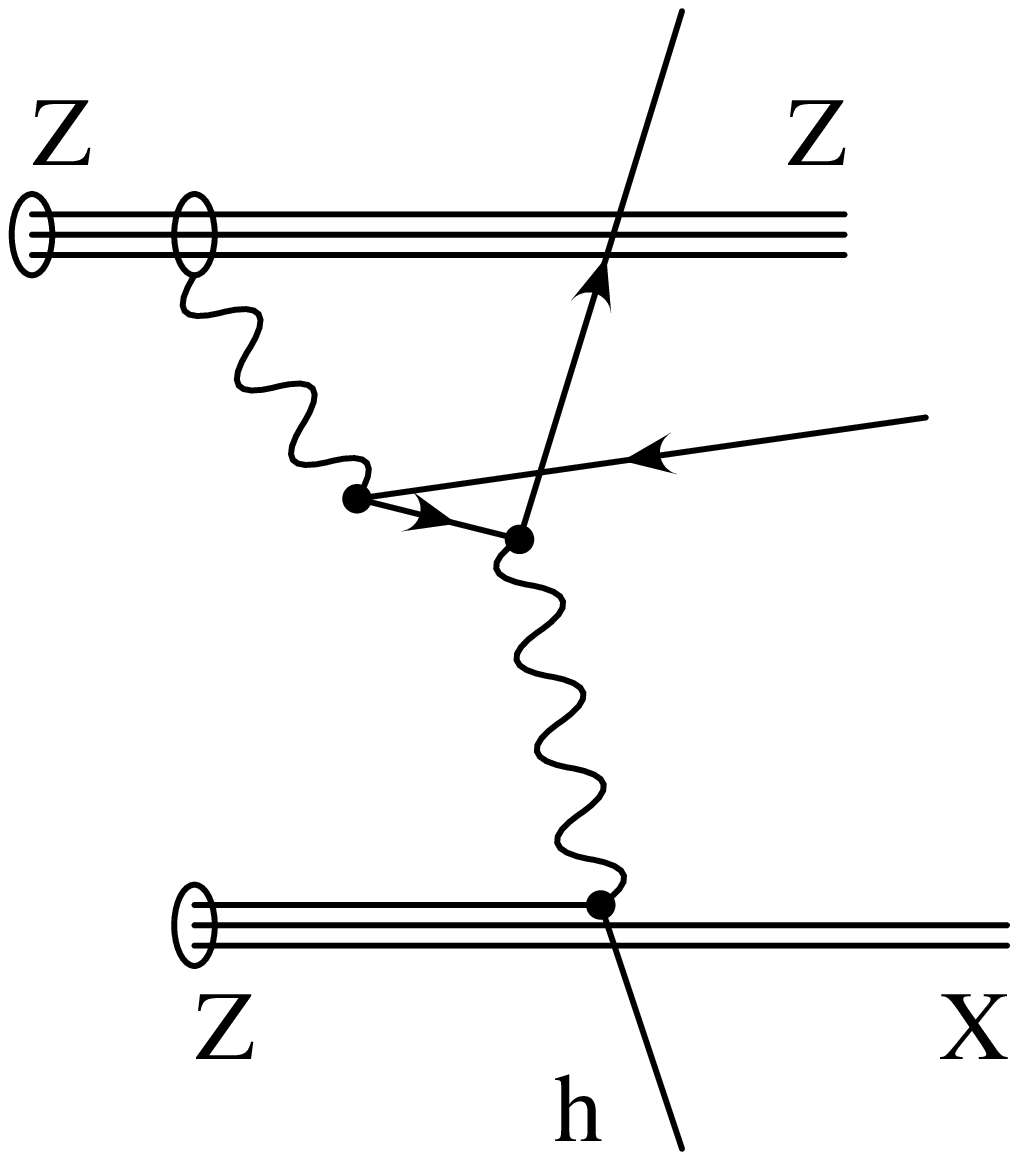}}(b)
  \end{center}
  \caption{Two possible options to measure the quark pdf of nuclei with
  UPCs: (a) Compton scattering on the quark with large transverse
  momenta, (b) inelastic pair production with one highly virtual photon.}
  \label{fig:EMA}
\end{figure}

In connection with the change of the gluon distribution functions also
a change of the quark distribution functions is expected. These
distribution functions are accessible at a large $Q^2$ scale, e.g., through
the Compton scattering process 
$\gamma q\rightarrow \gamma q$ at large transverse momenta \cite{Vogt:2004yr}.
An alternative approach for smaller 
$Q^2$ scale is to use inelastic pair production \cite{Dreyer04}, see
Fig.~\ref{fig:EMA}(b). Whereas one photon is emitted elastically, the
other is highly virtual. Therefore one can see this as lepton-ion 
deep inelastic scattering.
%A simple picture of this
%process is the ``equivalent lepton approximation'': The ions have
%photons as partons, which are again composed of leptons (electrons or
%muons) as their partons. Therefore in this way one studies lepton-ion
%deep inelastic scattering \cite{Dreyer04}.

\section{Summary and Conclusions}

Ultraperipheral collisions at the LHC allow to study photon-photon and
photo-nucleus processes at high energies and large luminosities. In
photon-photon physics electroweak processes can be studied in tagged
$pp$ collisions, meson production, double vector meson production are
clearly possible. The total hadronic cross section
$\gamma\gamma\rightarrow$~hadron would be an interesting study, but is
probably difficult to do. The
discovery potential for new physics seems unfortunately to be rather
limited.

Photon-nucleus collisions can be used to study coherent vector meson
production, especially of the $J/\Psi$ and probably also the
$\Upsilon$. Photon-gluon fusion allows to study gluon-pdfs in
nuclei. Quark distributions can be either studied by Compton
scattering of photons or inelastic pair production.

UPC are also of practical importance for ion beams at the LHC:
Bound-free pair production and electromagnetic excitation are the
dominant loss processes at Pb beams and also limit the maximum
achievable beam luminosity. Mutual electromagnetic excitation but also
lepton pair production are interesting possibilities for luminosity
measurements. The new measurements from RHIC help the LHC here.

At CERN a Yellow Report to document the physics
potential for UPC collisions is currently being
prepared. Details and also the talks of some workshops can be found
at their webpage \cite{quasarupc}.

To summarize, let us just say that {\it ``The events are there, some of
them are most interesting, just
do not throw them away''.}

\lastevenpage

\bigskip

\begin{thebibliography}{10}
\bibitem{BaurHTS02}
G.~Baur, K.~Hencken, D.~Trautmann, S.~Sadovsky and Y.~Kharlov,
{\em Phys.\ Rept.\ } {\bf 364}, 359 (2002)
[arXiv:hep-ph/0112211].

\bibitem{BaurHT98}
G.~Baur, K.~Hencken, and D.~Trautmann,
\newblock {\em Topical Review, J. Phys. G} 24 (1998) 1657.

\bibitem{KraussGS97}
F.~Krauss, M.~Greiner, and G.~Soff,
\newblock {\em Prog. Part. Nucl. Phys.} 39 (1997) 503.

\bibitem{BertulaniB88}
C.~A. Bertulani and G.~Baur,
\newblock {\em Phys. Rep.} 163 (1988) 299.

\bibitem{ALICEPPR} ALICE Collaboration, 
   {\it ALICE Physics Performance Report}, vol 1,
   CERN/LHCC/2003-049.

\bibitem{ATLASHI} Aronson {\it et al.}, ArXiv:nucl-ex/0212016.

\bibitem{CMSHI} {\em Heavy Ion Physics Programme in CMS.}
G. Baur {\it et al.}, CMS NOTE-2000/060, to appear also
in Eur. Phys. J. C.

\bibitem{CMS-REVIEW}Heavy-Ion Physics at the LHC with the Compact
Muon Solenoid Detector, 
http://yepes.rice.edu/cms/reviewJune2004/CmsHiReview.pdf

\bibitem{Fermi24}
E.~Fermi,
\newblock {\em Z. Phys.} 29 (1924) 315.

\bibitem{Williams34}
E.~J. Williams,
\newblock {\em Phys. Rev.} 45 (1934) 729.

\bibitem{BudnevGM75}
V.~M. Budnev {\it et~al.},
\newblock {\em Phys. Rep.} 15 (1975) 181.

\bibitem{BaltzS98}
A.~J. Baltz and M.~Strikman,
\newblock {\em Phys. Rev.~D} 57 (1998) 548.

\bibitem{BrandtEM94b}
D.~Brandt, K.~Eggert, and A.~Morsch,
\newblock CERN SL/94-04(AP), 1994.

\bibitem{EngelRR97}
R.~Engel {\it et~al.},
\newblock {\em Z. Phys. C} 74 (1997) 687.

\bibitem{Felix97}
K.~Eggert {\it et~al.},
FELIX Letter of Intent, CERN/LHCC 97--45, LHCC/I10

\bibitem{FELIX} A.~Ageev {\it et al.}, {\em J. Phys. G: Nucl. Part. Phys.}\ {\bf 28}, R117 (2002).

\bibitem{DreesGN94}
M.~Drees {\it et~al.},
\newblock {\em Phys. Rev.~D} 50 (1994) 2335.

\bibitem{OhnemusWZ94}
J. Ohnemus, T.~F. Walsh, and P.~M. Zerwas, {\em Phys. Lett.}~B {\bf 328},  369
  (1994).

\bibitem{BaurHTS98}
G.~Baur {\it et~al.}, CMS Note 1998/009.

\bibitem{Krawczyk96}
M.~Krawczyk,
\newblock in {\em Future Physics at HERA}, edited by G.~Ingelman, A.~{De
  Roeck}, and R.~Klanner, DESY, Hamburg, 1996,
\newblock IFT 21/96, [Arxiv:hep-ph/9609477].

\bibitem{ChoudhuryK97}
D.~Choudhury and M.~Krawczyk,
\newblock {\em Phys. Rev.~D} 55 (1997) 2774.

\bibitem{LiettiNRR01}
S.~M. Lietti {\it et~al.},
\newblock {\em Phys. Lett.~B} 497 (2001) 243.

\bibitem{HenckenKKS96}
K.~Hencken {\it et~al.},
\newblock TPHIC, event generator of two photon interactions in heavy ion
  collisions,
\newblock IHEP-96-38, 1996.

\bibitem{GinzburgS98}
I.~F. Ginzburg and A.~Schiller,
\newblock {\em Phys. Rev.~D} 57 (1998) R6599.

\bibitem{Abbott98}
B.~Abbott,
\newblock {\em Phys. Rev. Lett.} 81 (1998) 524.

\bibitem{RoldaoN00}
C.~G. Rold{\~a}o and A.~A. Natale,
\newblock {\em Phys. Rev.~C} 61 (2000) 064907.

\bibitem{Piotrzkowski00}
K.~Piotrzkowski,
\newblock {\em Phys. Rev.~D} 63 (2000) 071502.

\bibitem{Piotrzkowskihere}
K.~Piotrzkowski, Talk at this conference.

\bibitem{L3:97}
{L3 collaboration},
\newblock {\em Phys. Lett.~B} 408 (1997) 450.

\bibitem{Acciarri00b}
M.~Acciarri {\it et~al.},
\newblock {\em Phys. Lett.} B503 (2001) 10.

\bibitem{L3:01}
M.~Acciarri {\it et al.},
{\em Phys. Lett.} B519 (2001) 33..

\bibitem{Adams04}
J.~Adams {\it et al.},
arXiv:nucl-ex/0404012, to appear in Phys. Rev. C (2004).

\bibitem{HenckenBT04}
K.~Hencken, G.~Baur and D.~Trautmann,
{\em Phys.\ Rev.}\ C {\bf 69} (2004) 054902
[arXiv:nucl-th/0402061].

\bibitem{Klein01}
S.~R. Klein,
\newblock {\em Nucl. Instrum. Methods} A59 (2001) 51.

\bibitem{Brandt00}
D.~Brandt,
\newblock Review of the LHC Ion Report,
\newblock LHC Project Report 450, 2000.

\bibitem{BaronB93}
N.~Baron and G.~Baur,
\newblock {\em Phys. Rev.~C} 48 (1993) 1999.

\bibitem{VidovicGS93}
M.~Vidovi{\'c}, M.~Greiner, and G.~Soff,
\newblock {\em Phys. Rev.~C} 48 (1993) 2011.

\bibitem{Baltz98}
A.~Baltz, C.~Chasman, and S.~N. White,
\newblock {\em Nucl. Instrum. Methods} 417 (1998) 1.

\bibitem{Pshenichnov00}
I.~A. Pshenichnov {\it et~al.},
\newblock {\em Phys. Rev.~C} 57 (1998) 1920.

\bibitem{Pshenichnov01}
I.~A. Pshenichnov {\it et~al.}, {\em Phys.Rev.} C64 (2001) 024903

\bibitem{White01}
M.~Chiu {\it et~al.}, {\em Phys.Rev.Lett.} 89 (2002) 012302.

\bibitem{Pshenichnov98}
I.~A. Pshenichnov {\it et~al.},
\newblock {\em Phys. Rev.~C} 57 (1998) 1920.

\bibitem{Adler02}
C.~Adler {\it et al.}  [STAR Collaboration],
{\em Phys.\ Rev.\ Lett.}\  {\bf 89}, 272302 (2002);

\bibitem{KleinN99}
S.~Klein and J.~Nystrand,
\newblock {\em Phys. Rev.~C} 60 (1999) 014903.

\bibitem{KleinN00}
S.~Klein and J.~Nystrand,
\newblock {\em Phys. Rev. Lett.} 84 (2000) 2330.

\bibitem{Baltz:2002pp}
A.~J. Baltz, S.~R. Klein, and J. Nystrand, {\em Phys. Rev. Lett.} {\bf 89},  012301
  (2002).

\bibitem{cheunghere}
S.-U.~Chung, Talk at this conference.

\bibitem{Nystrandhere}
J.~Nystrand, Talk at this conference.

\bibitem{Klein04}
S.~R.~Klein  [STAR Collaboration],
arXiv:nucl-ex/0402007.

\bibitem{Frankfurt:2002wc}
L.~Frankfurt, M.~Strikman and M.~Zhalov,
{\em Phys.\ Lett.\ B} {\bf 537} (2002) 51
[arXiv:hep-ph/0204175].

\bibitem{fszstar}
L.~Frankfurt, M.~Strikman and M.~Zhalov,  {\em Phys.\ Rev.}\ {\bf C67}, 034901 (2003);

\bibitem{Baur:2003aa}
G. Baur {\it et al.}, {\em Nucl. Phys.} {\bf A729}, 787 (2003).

\bibitem{KleinNV00}
S.~R. Klein, J.~Nystrand, and R.~Vogt,
{\em Eur. Phys. J.} {\bf C21} (2001) 563.

\bibitem{Klein:2002wm}
S.~R.~Klein, J.~Nystrand and R.~Vogt, {\em Phys.\ Rev.}\ C {\bf 66} (2002) 044906

\bibitem{Vogt:2004yr}
R.~Vogt, arXiv:hep-ph/0407298.

\bibitem{Dreyer04} U. Dreyer {\it et al.}, to be submitted (2004).

\bibitem{quasarupc} See the homepage at http://quasar.unibas.ch/UPC/.

\end{thebibliography}
\end{document}